\documentclass[11pt]{article}

\newtheorem{theorem}{Theorem}
\newtheorem{lemma}{Lemma}
\newtheorem{corollary}{Corollary}

\newtheorem{proposition}{Proposition}

\newtheorem{observation}{Observation}

\newenvironment{proof}{\noindent\mbox{\bf Proof }}{\hfill\mbox{ 
$\bigtriangledown$}\medskip\par}

\begin{document}

\newcommand{\newspacing}{\baselineskip=1.2\normalbaselineskip}
\newspacing

\title{\vspace*{.1cm}\bf On the Importance of Having an Identity or, \\
is Consensus really Universal?}
\author{\small Harry Buhrman \\
\small CWI, Amsterdam
\and
\small Alessandro Panconesi \\
\small DSI, La Sapienza di Roma 
\and 
\small Riccardo Silvestri \\
\small DSI, La Sapienza di Roma 
\and
\small Paul Vitanyi \\
\small CWI, Amsterdam
}
\date{}
\maketitle

\begin{abstract}
We show that Naming-- the existence of distinct IDs known to all-- is
a hidden, but necessary, assumption of Herlihy's universality result
for Consensus. We then show in a very precise sense that Naming is harder than Consensus
and bring to the surface some important differences existing
between popular shared memory models.
\end{abstract}

\section{Introduction}
The consensus problem enjoys a well-deserved reputation in the
(theoretical) distributed computing community. 
Among others, a seminal paper of Herlihy added further evidence
in support of the claim that consensus is indeed a key theoretical construct 
\cite{herlihy}. Roughly speaking, Herlihy's paper
considers the following problem: Suppose that, besides
a shared memory, the hardware of our asynchronous, parallel machine
is equipped with objects (instantiations) of certain
abstract data types $T_1, T_2, \ldots, T_k$;
given this, is it possible to implement objects of a new abstract data
type $Y$ in a fault-tolerant manner?
The notion of fault-tolerance 
adopted here is that of wait-freedom, i.e. $(n-1)$-resiliency \cite{herlihy}. 
This question is the starting point of an interesting theory leading 
to many results and further intriguing questions
(see \cite{herlihy,jayanti} among others). One of the
basic results of this theory, already contained in the original article
of Herlihy, can be stated, somewhat loosely, as follows:
If an abstract data type $X$, together with a shared memory,
is powerful enough to
implement consensus for $n$ processes in a fault-tolerant manner then, $X$,
together with a shared memory, is also powerful enough to implement
in a fault-tolerant manner for $n$ processes
any other data structure $Y$. This is Herlihy's celebrated
universality result for consensus.

In this paper we perform an analysis of some of the basic
assumptions underlying Herlihy's result and discover several interesting
facts which, in view of the above, are somewhat counter-intuitive
and that could provocatively be summarized by the slogans
``consensus without naming is not universal'' and 
``naming with randomization is universal.''
To state our results precisely we shall recall some definitions
and known results.

In the {\bf consensus} problem we are given a set of 
$n$ asynchronous processes that, as far as this paper is concerned,
communicate via a shared-memory. 
Every process has its own input bit and is to produce its own output bit.
Processes can suffer from {\bf crash failures}.
The problem is to devise a protocol that can withstand up to
$(n-1)$ crash failures, i.e. a {\bf wait-free} protocol, satisfying the following
conditions:
\begin{itemize}
\item
Every non-faulty process terminates;
\item
All output bits are the same and,
\item
The output bit is the input bit of some process.
\end{itemize}

The {\bf naming} problem on the other hand, is as follows:
Devise a protocol for a set of $n$ asynchronous processes
such that, at the end, each non faulty process has selected
a unique identifier (key).
If processes have identifiers to start with then we have the
{\bf renaming} problem.  

In some sense, this paper is about the relative complexity of naming
to consensus, and viceversa.
We shall mostly concern ourselves with
{\bf probabilistic protocols}-- every process in the system,
modeled as an i/o automaton, has access to its own source of
unbiased random bits-- for systems consisting of
{\bf asynchronous} processes communicating via a {\bf shared memory}.
The availability of objects of abstract data type {\tt consensus}
and {\tt naming} is assumed. An object of type {\tt consensus} is a subroutine with input 
parameter $b \in \{0,1\}$. When invoked
by a process $p$ a bit $b'$ is returned. This bit is the same to all
invoking processes and is equal to some of the input bits, i.e. if $b'$ is returned some $p$
must have invoked the object with input parameter $b'$.
An object of type {\tt naming} is a subroutine without input parameters that,
when invoked by a process $p$, returns a value $v_p \in \{1,..,n\}$, $n$ being
the overall number of
processes. For any two processes $p \neq q$ we have that $v_p \neq v_q$.

The protocols we devise should be wait-free in spite of the {\bf adversary},
the ``malicious'' non-deterministic scheduling agent (algorithm) modeling
the environment. The adversary decides which, among
the currently pending operations, goes on next.  Pessimistically one assumes
that the adversary is actually trying to force the
protocol to work incorrectly and that the next scheduling decision--
which process moves next-- can be based on the whole past history of
the protocol execution so far.  This is the so-called {\em adaptive}
or {\bf strong} adversary.
In contrast, sometimes it is assumed that the adversary decides the
entire execution schedule beforehand. This is the so-called
{\em oblivious} or {\bf weak} adversary.

In the literature two shared-memory models are widespread.
The first assumes multiple reader - multiple writer registers.
In this model each location of the shared memory can be written
and read by any process. 
The other model assumes multiple reader - single writer
registers. Here, every register is owned by some unique process, which is
the only process that can write on that register, while every 
process is allowed to read the contents of any register.
In both models reads and writes are atomic operations; in case of
concurrent access to the same register it is assumed that the
adversary ``complies with'' some non-deterministic, but fair, policy. 
In this paper we shall refer to the first as the {\bf symmetric}
memory model and to the second as the {\bf asymmetric} memory model.

We are now ready to state the results of this paper.
Let us start by restating Herlihy's universality result
in our terminology. \newline

\noindent
{\bf Theorem.} [Herlihy]
{\em
Suppose that $n$ asynchronous processes interact via a shared memory
and that,
\begin{itemize}
\item[\em (i)]
Memory is symmetric;
\item[\em (ii)]
Each process has its own unique identifier;
\item[\em (iii)]
Objects of type {\tt consensus} are available to the processes.
\end{itemize}
Then, any abstract data type $T$ can be implemented in a wait-free
manner for the $n$ processes. } \newline

The first question we consider in this paper is: What happens
if the second hypothesis is removed? Can distinct identifiers
be generated from scratch in this memory model? 
The answer is negative, even assuming the availability of
consensus objects.
\begin{proposition}\label{namImp}
{\em [Naming is impossible]}
Suppose that $n$ asynchronous processes \underline{without identifiers}
interact via a shared memory and that,
\begin{itemize}
\item[\em (i)]
Memory is symmetric;
\item[\em (ii)]
Each process has access to its own source of unbiased random-bits;
\item[\em (iii)]
Objects of type {\tt consensus} are available to the processes;
\item[\em (iv)]
The adversary is weak.
\end{itemize}
Yet, wait-free Las Vegas naming is impossible. 
\end{proposition}
This result is simple to prove, but it is interesting in several respects.
First, it says that in a model more powerful than
Herlihy's, no protocol can produce distinct identifiers with certainty.
Therefore consensus by itself is {\bf not} universal, 
for wait-free naming objects cannot be implemented in a wait-free manner
with consensus alone. 

Recall that a Las Vegas protocol is always correct and that only
the running time is a random variable, while for a
Montecarlo protocol correctness too is a random variable.
Note that Montecarlo naming is trivial-- each process generates
$O(\log n)$ many random bits and with probability $1-o(1)$ no two
of them will be identical. Therefore, at least at the outset, only the question
of the existence of Las Vegas protocols is of interest.

Proposition~\ref{namImp} shows that the power of randomization to
``break the symmetry''is limited. If we start from a completely
symmetric situation, it is
impossible to generate identifiers that are surely distinct.

In stark contrast with the previous result, as we prove in this paper, the following
holds.
\begin{theorem}\label{easyCon}
{\em [Consensus is easy]}
Suppose that $n$ asynchronous processes \underline{without identifiers}
interact via a shared memory and that,
\begin{itemize}
\item[\em (i)]
Memory is symmetric;
\item[\em (ii)]
Each process has access to its own source of unbiased random-bits;
\item[\em (iii)]
The adversary is strong.
\end{itemize}
Then, there exist Las Vegas, wait-free consensus protocols for $n$
processes whose complexity is polynomial in expectation
and with high probability.
\end{theorem}
Notice that while Proposition~\ref{namImp} establishes the impossibility
of naming even against the weak adversary, here 
the adversary is strong.

Incidentally, Theorem~\ref{easyCon} shows that hypothesis~(iii) of
Proposition~\ref{namImp} is superfluos, for consensus objects
can be simulated via software in a wait-free manner.
It is well-known that hypothesis~(ii) is necessary, even if the adversary is weak
(see, for instance, \cite{lynch,aw}).

In some sense naming captures the notion of complete asymmetry among the processes and
if we start from a completely symmetric situation it embodies the intuitive notion of
complete break of symmetry.
It is well-known that randomization is a powerful ``symmetry-breaker'' although, as
Proposition~\ref{namImp} shows, not enough for naming, if we start
from a perfectly symmetric situation. 
This leads to the question of ``how much'' asymmetry is needed for naming. 
Let us consider therefore asymmetric memory, assuming moreover
that processes do not have 
access to the address (index) of their private registers. Formally, 
each process $p$ accesses the $m$ registers by means of a permutation
$\pi_p$. Register $\pi_p^i$-- $p$'s $i$th register--
will always be the same register,
but, for $p \neq q$,  $\pi_p^i$ and  $\pi_q^i$ might very well differ.
In particular processes cannot obtain the physical address of the
registers.  therefore the asymmetry is somehow hidden from the processes.
Although our motivation 
is mainly theoretical, this model has been used to study 
certain situations
in large dynamically changing systems where a consistent
indexing scheme 
is difficult or impossible to maintain \cite{LP90}. Moreover
this model could make sense in
cryptographical systems where this kind of consistency is to be avoided.

We show the following.
If the memory is initialized to all 0's (or any other fixed value)
we say that it is initialized {\em fairly}.
\begin{proposition}\label{namImpAsym}
Assume that the memory is initialized fairly.
If processes are identical, deterministic i/o automata without
identifiers then naming is impossible, even if memory is asymmetric
and the adversary is weak.
\end{proposition}
Thus, by themselves, neither randomization nor asymmetric memory can break the symmetry.
What is needed is their acting together.
\begin{theorem}\label{asymMem}
Suppose that $n$ asynchronous processes \underline{without identifiers}
interact via a shared memory and that,
\begin{itemize}
\item[\em (i)]
Memory is asymmetric and initialized fairly;
\item[\em (ii)]
Each process has access to its own source of unbiased random-bits;
\item[\em (iii)]
The adversary is strong.
\end{itemize}
Then, there exist a Las Vegas, wait-free naming protocol for $n$
processes whose running time is polynomial in expectation.
Furthermore the key space from which identifiers
are drawn has size $n$, which is optimal.
\end{theorem}
Therefore, with randomization, asymmetric memory is inherently symmetry-breaking,
whereas consensus is not.

This result improves on previous work in \cite{pptv} in which 
naming protocols with almost-optimal key range are given.
We prove two versions of the above result. We first give a simple protocol whose running time
is $\Theta(n^2 \log n)$ w.h.p. and a faster protocol, named {\tt squeeze},
whose expected running time is $O(n \log^3 n)$. 
As a by-product we also show that an object we call {\tt selectWinner}
cannot be implemented by consensus alone, i.e. without naming, even if randomization is available 
and the adversary is weak. The semantics of {\tt selectWinner} is the following:
the object selects a unique winner among the invoking processes. 

Since any deterministic
protocol must use a key range of size at least $2n-1$ in order to be wait-free
\cite{HerS93}, this is yet another instance in which
randomization is more powerful than determinism 
as far as fault-tolerant computing is concerned.

Our results show, perhaps surprisingly, that 
multiple reader - single writer registers are more powerful than
multiple reader - multiple writer registers, even though the latter
might represent a faster alternative.
This highlights an important difference  
between the two models. 

Our Theorem~\ref{easyCon} is obtained by combining several
known ideas and protocols, in particular those in \cite{aspnes-herlihy} and
\cite{chandra}. When compared to the protocol in \cite{aspnes-herlihy} it is, we believe,
simpler, and its correctness is easier
to establish (see, for instance, \cite{segala}). Moreover, it works in the less
powerful symmetric model and
can deal with the strong adversary, whereas the protocol in \cite{chandra} can only withstand
the ``intermediate'' adversary, whose power lies somewhere between the more traditional
weak and strong adversaries we consider.
From the technical point of view, our Propositions~\ref{namImp} and~\ref{namImpAsym}
are essentially contained in \cite{kops} to which we refer for other interesting related results.  
Other related work can be found in \cite{agm, jt}.

In spite of the fact that we make use of several known technical ingredients,
our analysis, we believe, is novel and brings to light for the
first time new and, we hope, interesting aspects of fundamental concepts.

\newcommand{\isp}{\hspace{0.7cm}}
\newcommand{\Mark}[2]{\mbox{{\sc Mark}{\rm [}$#1$, $#2${\rm ]}}}
\newcommand{\Coin}[2]{\mbox{{\sc GetCoin}$_{\delta}${\rm (}$#1$, $#2${\rm )}}}
\newcommand{\lr}{$\;\leftarrow\;$}
\newcommand{\estp}{\mbox{$myTeam_p$}}
\newcommand{\otherTeam}{\mbox{$otherTeam_p$}}
\newcommand{\nestp}{\mbox{$tentativeNewTeam_p$}}
\newcommand{\posp}{\mbox{$position_p$}}

\section{Consensus is Easy, Naming is Hard}
We start by outlining a consensus protocol
assuming that (a) the memory is symmetric, (b) processes are
i/o automata {\bf without} identifiers which have access to
their own source of (c) random bits. 
Our protocol is obtained by combining together several known ideas
and by adapting them to our setting.
The protocol, a randomized implementation of $n$-process binary
consensus for symmetric memory, is a modification of the protocol
proposed by Chandra \cite{chandra}. The original protocol cannot be used in
our setting since its shared coins require that processes have
unique IDs. Thus, we combine it with a modification of the weak
shared coin protocol of Aspnes and Herlihy \cite{aspnes-herlihy}. The latter
cannot be directly used in our setting either, since it requires
asymmetric memory. Another difference is that, unlike in Chandra's protocol,
we cannot revert to Aspnes' consensus \cite{aspnes}. 
In this paper we are only interested in establishing the existence of a
polynomial protocol and make no attempt at optimization.
Since the expected running time of our protocol is polynomial, by Markov's Inequality,
it follows that the running time and, consequently, the space used are
polynomial with high probability (inverse polynomial probability of failure).
Conceivably superpolynomial space could be needed. We leave it as an open
problem whether this is necessary.
In the sequel we will assume familiarity with the notion
of weak shared coin of \cite{aspnes-herlihy} to which the reader is referred.

The protocol, shown in Figure~\ref{fig:chandra}, is 
based on the following idea. Processes engage in a race of sorts by splitting
into two groups: those supporting the 0 value and those supporting the
1 value. At the beginning membership in the two ``teams'' is decided
by the input bits. Corresponding to each team there is a ``counter'',
implemented with a row of contiguous ``flags''-- the array of booleans \Mark{}{}-- which are to be raised 
one after the other from left to right by the team members, cooperatively and asynchronously. 
The variable \posp\ of each process $p$ records
the rightmost (raised) flag of its team the process knows about. 
The protocol keeps executing the following loop, until a decision is
made. The current team of  process $p$ is defined by the variable \estp.
The process first increments its own team counter by
raising the \posp-th flag of its own team (this might have already been
done by some other team member, but never mind). For instance, if $p$'s team corresponds to
the value $b$ then, \Mark{\posp}{b} is set to true.
Thus, as far as process $p$ is concerned, the value of
its own team counter is $position_p$ (of course, this might not 
accurately reflect the real situation).  The process then
``reads'' the other counter by looking at the other team's row
of flags at positions $position_p+1, position_p, position_p-1$, in this order.
There are four cases to consider:
(a) if the other team is ahead the process sets the variable \nestp\
to the other team;
(b) if the two counters are equal, the process flips a fair coin
$X \in \{0,1\}$ by invoking the protocol \Coin{}{}  and sets
\nestp\ to $X$; 
(c) if the other team trails by one, the process sticks to its team, and 
(d) if the other team trails by two (or more) the process decides
on its own team and stops executing the protocol. 
The setting of \nestp\ is, as the name suggests, tentative.
Before executing the next iteration, the process checks again
the counter of its own team. If this has been changed in the meanwhile
(i.e. if the $({\posp}+1)$-st flag has been raised)
then the process sticks to his old team and continues; otherwise, it does
join the team specified by \nestp. 
The array \Mark is implemented with {\tt multiple reader -
multiple writer} registers, while the other variables are local to each
process and accessible to it only. The local variables can assume only a finite (constant) set of
values and can therefore be ``hardwired'' in the states of the i/o automaton
representing the process. 

The only, but crucial, difference between our protocol and that of
Chandra concerns procedure \Coin{}{}.
In Chandra's setting essentially it is possible to implement
``via software'' a {\em global coin}, thanks to the naming assumption
and the special assumption concerning the power of the adversary (``intermediate'' instead of
strong).
In the implementation in Figure~\ref{fig:chandra}, we use a protocol
for a weak shared coin for symmetric memory. For every 
$b\in \{0,1\}$ and every $i \geq 1$ an independent realization of the
weak shared coin protocol is performed. An invocation of such a
protocol is denoted by \Coin{b}{i}, where $\delta$ is a positive real
that represents the agreement parameter of the weak shared coin
(see \cite{aspnes-herlihy}).  \Coin{b}{i} satisfies th efollowing conditions.
Upon invocations with values $b$ and $i$, it returns 0 to all invoking
processes with probability $p \geq (1-\delta)/2$; it returns 1 to all invoking
processes with probability $p \geq (1-\delta)/2$; and, it returns 0 to some
and 1 to the others with probaility at most $\delta$ \cite{aspnes-herlihy}.

\begin{figure}
\begin{center}
\hrule
\small
\begin{tabular}{ll}
\\
\multicolumn{2}{l}{{\tt \{}{\em Initialization}{\tt \}}}\\
\multicolumn{2}{l}{\Mark{0}{0}, \Mark{1}{0} \lr {\tt true}}\\
& \\
\multicolumn{2}{l}{{\tt \{}{\em Algorithm for process $p$}{\tt \}}}\\
& \\
\multicolumn{2}{l}{{\bf function} {\em propose}($v$): {\bf return}s 0 or 1}\\
1. & \estp \lr $v$; \otherTeam \lr 1 - \estp\\
2. & \posp \lr 1\\
3. & {\bf repeat}\\
4. & \isp \Mark{\estp}{\posp} \lr {\tt true}\\
5. & \isp {\bf if} \Mark{\otherTeam}{\posp + 1}\\
6. & \isp\isp \nestp \lr $1 - \estp$\\
7. & \isp {\bf else if} \Mark{\otherTeam}{\posp}\\
8. & \isp\isp \nestp \lr \Coin{\estp}{\posp}\\
9. & \isp {\bf else if} \Mark{\otherTeam}{\posp - 1}\\
10. & \isp\isp \nestp \lr \estp\\
11. & \isp {\bf else return}(\estp) {\tt \{}{\em Decide} \estp{\tt \}}\\
12. & \isp {\bf if not} \Mark{\estp}{\posp + 1}\\
13. & \isp\isp \estp \lr \nestp\\
14. & \isp \posp \lr $\posp + 1$\\
    & {\bf end repeat} \\
\\
\end{tabular}
\hrule
\end{center}
\caption{$n$-process binary consensus for symmetric memory}
\label{fig:chandra}
\end{figure}

First, we prove that the protocol in Figure~\ref{fig:chandra} is correct
and efficient. Later we show how to implement the weak shared coin.

\begin{lemma}\label{sym-cons-valid}
If some process decides $v$ at time $t$, then, before time $t$ some process started
executing propose($v$).
\end{lemma}

\begin{proof}
The proof is exactly the same of that of Lemma~1 in \cite{chandra}. 
\end{proof}

\begin{lemma}\label{sym-cons-consistent}
No two processes decide different values.
\end{lemma}

\begin{proof}
The proof is exactly the same of that of case (3) of Lemma~4 in \cite{chandra}.
\end{proof}

\begin{lemma}\label{sym-cons-shared}
Suppose that the following conditions hold:
\begin{description}
\item[{\em i)}] \Mark{b}{i} $=$ {\tt true} at time $t$,
\item[{\em ii)}] \Mark{1 - b}{i} $=$ {\tt false} before time $t$,
\item[{\em iii)}] \Mark{1 - b}{i} is set {\tt true} at time $t'$ ($t' > t$),
and
\item[{\em iv)}] every invocation of both \Coin{b}{i} and 
\Coin{1 - b}{i} yields value $b$.
\end{description}
Then, no process sets \Mark{1 - b}{i + 1} to {\tt true}.
\end{lemma}

\begin{proof}
The proof is essentially the same of that of the Claim included in
the proof of Lemma 6 in \cite{chandra}.
\end{proof}

The next lemma is the heart of the new proof. The difficulty of course is that now
we are using protocol \Coin{}{} instead of the ``global coins'' of \cite{chandra}, and have
to contend with the strong adversary. 
The crucial observation is that if two teams are in the same position $i$ and the adversary
wants to preserve parity between them, it must allow both teams to raise
their flags ``simultaneously,'' i.e. at least one teammate in each team must observe
parity in the row of flags. But then each team will proceed to invoke
\Coin{}{}, whose unknown outcome is unfavorable to the adversary 
with probability at least $(\delta/2)^2$.

\begin{lemma}\label{sym-cons-prob}
If \Mark{b}{i} $=$ {\tt true} at time $t$ and \Mark{1 - b}{i} $=$
{\tt false} before time $t$, then with probability at least $\delta^2/4$,
\Mark{1 - b}{i + 1} is always {\tt false}.
\end{lemma}

\begin{proof}
If \Mark{1 - b}{i} is always {\tt false}, then it can be shown that
\Mark{1 - b}{i + 1} is always {\tt false} (the proof is the same of that
of Lemma~2 in \cite{chandra}). So, assume that \Mark{1 - b}{i} is set to
{\tt true} at some time $t'$ (clearly, $t' > t$). Since no invocation
of both \Coin{b}{i} and \Coin{1 - b}{i} is made before time $t$, the
values yielded by these invocations are independent of the schedule
until time $t$. Thus, with probability at least $\delta^2/4$, 
all the invocations of \Coin{b}{i} and \Coin{1 - b}{i} yield
the same value $b$. From Lemma~\ref{sym-cons-shared}, it follows
that, with probability at least $\delta^2/4$, \Mark{1 - b}{i + 1} is 
always {\tt false}.
\end{proof}

\begin{theorem}\label{sym-cons-th}
The protocol of Figure~\ref{fig:chandra} is a randomized solution
to $n$-process binary consensus. Assuming that each invocation of \Coin{}{} 
costs one unit of time, the expected running time per process
$O(1)$. Furthermore, with high probability every process will invoke
\Coin{}{}  $O(\log n)$ many times.
\end{theorem}
\begin{proof}
From Lemma~\ref{sym-cons-consistent}, if any two processes decide, they
decide on the same value.  From Lemma~\ref{sym-cons-valid} we know that the
decision value is the input bit of some process. We now show that all processes
decide within a finite number of steps and that this number is polynomial both
in expectation and with high probability.

As regarding the expected decision time for any process, let $P(i)$ denote
the probability that there is a value $b \in \{0, 1\}$ such that
\Mark{b}{i} is always {\tt false}. From Lemma~\ref{sym-cons-prob},
it follows that
\[ P(i) \geq 1 - (1 - \delta^2/4)^{i - 1} \quad\quad i \geq 1 \]
Also, if \Mark{b}{i} is always {\tt false}, it is easy to see that
all the processes decide within $i + 1$ iterations of the {\bf repeat} 
loop. Thus, with probability at least $1 - (1 - \delta^2/4)^{i - 1}$,
all the processes decide within $i+1$ iterations of the {\bf repeat}
loop. This implies that the expected running time per process is
{\rm O(1)}. The high probability claim follows from the observation that pessimistically
the process describing the invocations of \Coin{}{} can be modeled as a geometric distribution
with parameter $p := (\delta/2)^2$.
\end{proof}

We now come to
the implementation of the weak shared coin for
symmetric memory, 
which we accomplish via a slight modification of the
protocol of Aspnes and Herlihy \cite{aspnes-herlihy}.
In that protocol the $n$ processes cooperatively simulate a random walk
with absorbing barriers.
To keep track of the pebble a {\tt distributed counter} is employed. The distributed counter
is implemented with an array of $n$ registers, with position $i$ privately owned
by process $i$ (that is, naming or asymmetric memory is assumed). When process $i$ wants
to move the pebble it updates atomically its own private register
by incrementing or decrementing it by one. The private register
also records another piece of information namely, the number
of times that the owner updated it (this allows one to show that the implementation of the read is
linearizable). On the other hand, reading the position of
the pebble is a non-atomic operation. To read the counter the process scans
the array of registers twice; if the two scans yield identical values the read is completed,
otherwise two more scans are performed, and so on.
As shown in \cite{aspnes-herlihy}, the expected number of elementary operations
(read's and write's) performed by each process is $O(n^4)$.

Since in our setting we cannot
use single-writer registers, we use an array {\sc C}[] of $n^2$ 
multiple-writer multiple-reader registers for the counter. The algorithm for
a process $p$ is as follows. Firstly, $p$ chooses uniformly at random
one of the $n^2$ registers of {\sc C}[], let it be the $k$th. Then,
the process proceeds with the protocol of Aspnes and Herlihy 
by using {\sc C}[$k$] as its own register and by applying the counting
operations to all the registers of {\sc C}[]. Since we are using $n^2$ registers
instead of $n$, the expected number of steps that each process performs
to simulate the protocol is  $O(n^5)$. The agreement parameter
of the protocol is set to $2e\delta$.
Since the expected number of rounds of the original protocol is {\rm O($n^4$)},
by Markov's Inequality, there is a
constant $B$ such that, with probability at least $1/2$, the protocol
terminates within $Bn^5$ rounds. It is easy to see
that if no two processes choose the same register,
then the protocol implements a weak shared coin with the same agreement
parameter of the original protocol in $O(n^5)$ many steps. 
To ensure that our protocol will terminate in any case, if after $Bn^5$ steps the process has not yet
decided then it flips a coin and decides accordingly. Thus, in any case
the protocol terminates returning a value $0$ or $1$ to the calling process
within $O(n^5)$ steps. The probability that 
no two processes choose the same register is 
\[ \left(1 - \frac{1}{n^2}\right)\left(1 - \frac{2}{n^2}\right)\cdots 
\left(1 - \frac{n - 1}{n^2}\right)
 \geq \frac{1}{e}. \]
Thus, the agreement parameter of our protocol is at least $1/2\cdot 1/e
\cdot 2e\delta = \delta$.  We have proved the following fact.
\begin{lemma}
For any $\delta > 0$, a weak shared coin with agreement parameter
$\delta$ can be implemented in the symmetric model (with randomization)
in $O(n^5)$ steps, even against the strong adversary. 
\end{lemma}
\begin{corollary}
The expected running time per process of the protocol of
Theorem~\ref{sym-cons-th} is $O(n^5)$.
\end{corollary}
\bigskip

We show next that, in contrast, no protocol exists in the symmetric model for naming, even assuming
the availability of consensus objects and the weak adversary.
\begin{theorem}\label{fact:namImp}
Suppose that an asynchronous, shared memory machine is such that:
\begin{itemize}
\item
the memory is symmetric;
\item
every process has access to a source of independent, unbiased random bits,
and
\item
consensus objects are available.
\end{itemize} 
Then, still, naming is impossible even against a weak adversary.
\end{theorem}
\begin{proof}
By contradiction suppose there exist such a protocol.
Consider two processes P and Q and let only Q go. Since the protocol is wait-free
there exists a sequence of steps $\sigma=s_1 s_2 \ldots s_n $ taken by Q
such that Q decides on a name 
$k_{\sigma}$. The memory goes through a sequence of states $m_0 m_1 \ldots m_n$.
The sequence $\sigma$ has a certain probability $p_{\sigma} = p_1 p_2 \ldots p_n$ of being
executed by Q. Start the system again, this time making both P and Q move, but one step at a time
alternating between P and Q. With probability $p_1^2$ both P and Q will make the same step $s_1$.
A simple case analysis performed on the atomic operations (read, write, invoke consensus)
shows that thereafter P and Q are in the same state and the shared memory
is in the same state $m_1$ in which it was when Q executed $s_1$ alone. This happens with probability 
$p_1^2$. With probability $p_2^2$, if P and Q make one more step each, we reach a situation
in which P and Q are in the same state and the memory state is $m_2$.
And so on, until, with probability
$p_{\sigma}^2$ both P and Q decide on the same identifier, a contradiction.  
\end{proof}

Thus,  naming is a necessary assumption in Herlihy's universality construction.


%

\section{Naming with Asymmetric Memory}
We now come to the question of whether asymmetric memory can be used to
break the symmetry. First we show that by itself 
it is not sufficient. Then we show that together with randomness
it allows naming to be solved in polynomial-time, using a key space of optimal size.

\begin{proposition} 
Suppose the memory is initialized fairly that is, all registers are initially set to 0
(or any other fixed value). Then,
if processes are identical, deterministic i/o automata without 
identifiers, naming is impossible, even if memory is asymmetric 
and the adversary is weak.
\end{proposition} 

\begin{proof}
Consider two processes $p$ and $q$ that are identical deterministic 
i/o automata without identifiers. The shared memory is asymmetric:
processes $p$ and $q$ access the $n$ registers by means of 
permutations $\pi_p$ and $\pi_q$. That is, when processor $p$ ($q$) 
performs a {\tt Read($i$)} operation, the result will be the content 
of the register having absolute index $\pi_p(i)$ ($\pi_q(i)$).
Analogously for the {\tt Write($i$, $v$)} operations. We assume that {\tt 
Write($i$, $v$)} is legal only if $i \leq n/2$. That is, the local indices 
of the private registers are $1,2,\ldots,n/2$.
We show the impossibility of naming even against the very simple
adversary with the alternating execution schedule: 
$p$, $q$, $p$, $q$, $\ldots$
In this schedule the execution proceeds in rounds. Each round 
consists of a step of $p$ followed by a step of $q$.

We need some notions.
We call the map of the contents of all the registers, shared and 
local, the absolute view. It is a function that maps each absolute 
index $i$ to the content of the register of absolute index $i$.
Given an absolute view $V$ remains determined 
the local views $L_p(V)$ and $L_q(V)$. The local view $L_p(V)$
is the map of the contents of the local registers of $p$ and 
of the shared registers through the permutation $\pi_p$. 
In other words, $L_p(V)$ is a function
that maps each local index $j$ to the content
of the register having local index $j$ w.r.t. the processor $p$. 
In particular, if $j$ is a local 
index of a shared register then the local view maps $j$ to the
content of the register of absolute index $\pi_p(j)$.
Analogously for the local view $L_q(V)$. We have already assumed that
the set of local indices of shared registers is 
the same for both processors (i.e. the set $\{1,2,\ldots,n\}$).
We further assume that the set of local indices of
local registers is the same for both processors. Thus, the entire
set of local indices is the same for both processors. Therefore,
the domains of the two local views coincide.
An instant configuration, or simply a configuration, for processor 
$p$ ($q$) is a pair $(L, s)$ where $L$ is a local view and $s$ 
is a state of $p$ ($q$).

Let $s^p_t$ and $s^q_t$ be the states of $p$ and $q$, respectively,
at the beginning of round $t$.
Let $V_t$ be the absolute view at the 
beginning of round $t$. We show that, for any $t$, if 
the configurations $(L_p(V_t), s^p_t)$ and $(L_q(V_t), s^q_t)$ 
are equal then after the execution of round $t$ the resulting
configurations $(L_p(V_{t+1}), s^p_{t+1})$ and $(L_q(V_{t+1}), s^q_{t+1})$ 
are equal again.
This fact together with the assumption that the initial 
configurations $(L_p(V_0), s^p_0)$ and $(L_q(V_0), s^q_0)$
are equal imply that the processes cannot select unique identifiers.

Suppose that $(L_p(V_t), s^p_t)$ and $(L_q(V_t), s^q_t)$ are equal.
Since $s^p_t = s^q_t$ and the processes are identical deterministic 
i/o automata, both processes execute the same operation {\tt op} at round $t$.
The operation {\tt op} can be a {\tt Read}, a {\tt Write}, or a local 
operation. 

If {\tt op} is a {\tt Read($i$)} then, since $L_p(V_t)$ and $L_q(V_t)$ 
are equal, the processors read from different registers that have 
equal contents. So, the changes to the local 
registers are the same and the local views $L_p(V_{t+1})$ and $L_q(V_{t+1})$
are equal. Consequently, also the states $s^p_{t+1}$ and $s^q_{t+1}$
coincide.

If {\tt op} is a {\tt Write($i$, $x$)} then, both processors write the same 
value to different registers. But these registers have the same local 
index. So, the next local views $L_p(V_{t+1})$ and $L_q(V_{t+1})$ are
equal again. Consequently, also the states $s^p_{t+1}$ and $s^q_{t+1}$
coincide.

If {\tt op} is a local operation then, since {\tt op} and the local 
memory is the same for both processors, the changes to the local 
memories are the same and the local views $L_p(V_{t+1})$ and $L_q(V_{t+1})$ are
equal. Consequently, also the states $s^p_{t+1}$ and $s^q_{t+1}$
coincide.
\end{proof}

Let us now turn to our naming protocol {\tt squeeze}. 
For now, let us assume the availability of
objects {\tt selectWinner(i)} with the following semantics.
The object is invoked with a parameter $i$; the response is
to return the value {\em ``You own key $i$!''} to exactly one of the invoking processes,
and {\em ``Sorry, look for another key''} to all remaining processes. The choice of the ``winner''
is non-deterministic. Later we will show that {\tt selectWinner} admits
a wait-free, polynomial time Las Vegas solution
in our setting. With {\tt selectWinner} a naming protocol can be easily obtained
as follows: Try each key one by one, in sequence, each time invoking {\tt selectWinner}.
This protocol, dubbed {\tt simpleButExpensive}, is shown in Figure~\ref{fig:easyNaming}. 
Therefore we obtain the following.
\begin{proposition}
Suppose that $n$ asynchronous processes \underline{without identifiers}
interact via a shared memory and that,
\begin{itemize}
\item[\em (i)]
Memory is asymmetric;
\item[\em (ii)]
Each process has access to its own source of unbiased random-bits;
\item[\em (iii)]
The adversary is strong.
\end{itemize}
Then, protocol {\tt simpleButExpensive} is a wait-free Las Vegas solution
to naming whose running time is polynomial in expectation and with high probability
\end{proposition}
The high probability statement follows from Lemma~\ref{lemma:selWinIter}.

Although the overall running time of {\tt simpleButExpensive} is polynomial, given the high
cost of invoking {\tt selectWinner} 
we turn our attention to protocol {\tt squeeze} which, in expecatation,
will only perform $O(\log^2 n)$ such invocations instead of linearly many.
\begin{figure}
\begin{center}
\small
\hrule
\tt
\begin{tabular}{l}
\\
protocol simpleButExpensive(): key; \\[.2cm]
begin \\ 
\ \ for k := 1 to n do  \\
\ \ \ \ if selectWinner(k) = {\em ``You own key $k$!''} then return(k); \\
end \\
\\
\end{tabular}
\hrule
\end{center}
\caption{Simple but expensive protocol for naming}\label{fig:easyNaming}
\end{figure}

In protocol {\tt squeeze}
the name space is divided into {\em segments}, 
defined by the following recurrence, where $p$ is 
a parameter between 0 and 1 to be fixed later: 
$$
s_{k} = p (1-p)^{k-1} n 
$$
To simplify the presentation we assume without loss of generality that all $s_i$'s are integral.
$s_{\ell}$ is the last value $s_i$ such that $s_i \ge \log^2 n$.
The first segment consists of the key interval $I_1 := [0,s_1)$;
the second segment consists
of the key interval $I_2 := [s_1,s_1+s_2)$; the third of the key interval 
$I_3 := [s_1+s_2,s_1+s_2+s_3)$, and so on. The final segment $I_{\ell+1}$ consists
of the last $n - \sum_{j=1}^{\ell} s_j$ keys. 
In the protocol, each process $p$ starts by selecting a tentative key
$i$ uniformly at random in $I_1$.
Then, it invokes ${\tt selectWinner(i)}$; 
if $p$ ``wins,'' the key becomes final and $p$ stops; otherwise,
$p$ selects a second tentative key $j$ uniformly at random in $I_2$. Again, 
${\tt selectWinner(j)}$ is invoked and if $p$ ``wins'' $j$ becomes final and $p$ stops, otherwise 
$p$ continues in this fashion until $I_{\ell+1}$ is reached. 
The keys of $I_{\ell+1}$ are tried one by one
in sequence. If at the end $p$ has no key yet, it will execute the
protocol {\tt simpleButExpensive} of Figure~\ref{fig:easyNaming} as a
back-up procedure.
The resulting protocol appears in Figure~\ref{fig:squeeze}.

Assuming the availability of objects of type {\tt selectWinner}, protocol
{\tt squeeze} assigns a key to every non-faulty process with probability 1. This 
follows, because the protocol ensures that {\tt selectWinner(i)} is invoked for every 
$i$, $1 \le i \le n$,
and each such invocation assigns a key to exactly one process.
We will now argue that with high probability every process receives a unique key
before the back-up procedure, and that therefore the number of invocations of
{\tt selectWinner} objects is $O(\log^2 n)$ per process w.h.p..

Protocol {\tt squeeze} maintains the following invariant (w.h.p.).
Let $P_k$ be the set of processes that after $k-1$ attempts still have to grab a key.
Their $k$-th attempt will be to select a key at random in segment $I_k$. Then, 
$|P_k| \approx |I_k| \log n \gg |I_k|$ (hence the protocol ``squeezes'' $P_k$ 
into $I_k$). Once the numbers are plugged in it follows that, with high probability,
every key in $I_k$ will be claimed by some process, and this for all $k$. Since
every key is claimed w.h.p. before the back-up procedure, every process, w.h.p.,
receives a key within $O(\log^2 n)$ invocations of {\tt selectWinner}.
By setting the parameter $p$ appropriately it is possible to keep 
the number of segments small, i.e. $O(\log^2 n)$, while maintaining
the invariant $|P_k| \gg |I_k|$ for each segment.

Let us focus first on a run without crashes.
Let $p_i$ be defined by the following recurrence
$$
p_i := (1-p)^{i-1} n.
$$
Clearly, $|P_i| \ge p_i$.
We want to show that, with high probability, $|P_i| = p_i$, for $i < \ell$.
If we can show this we are done because then $p_{\ell} = s_{\ell}$ and
the protocol ensures that every one of the remaining $p_{\ell}$ process
will receive one of the last $s_{\ell}$ keys.
A key $k$ is {\em claimed} if {\tt selectWinner(k)} is invoked by some process. 
Then, since there are no crashes,
$$
\Pr[\mbox{$\exists k \in I_i$, $k$ not claimed}] 
\le  \left(1 - \frac{1}{s_i} \right)^{p_i} 
\le  \exp \left\{- \frac{s_i}{p_i} \right\} 
\le  \exp \left\{- \frac{1}{p} \right\} 
= \frac{1}{n^c}
$$
for any fixed $c > 0$, provided that
$$
p := \frac{1}{c \log n}.
$$
With this choce of $p$ the number of segments $\ell$ is $O(\log^2 n)$.
The expected running time is therefore 
$$
E[T(n)] = O(\log^2 n) (1-\ell/n^{c}) + O(n) \ell/n^{c} = O(\log^2 n)
$$
for $c > 3$.

We now argue that with crashes the situation can only improve.
Let $C_1$ be the set of processes that crash before obtaining 
a response to their invocation of {\tt selectWinner} in $I_1$.
Let $F_1$ (F as in free) be the set of keys of $I_1$ that are not
claimed after processes in $P_1$ have made their
random choice. If $F_1$ is non empty, assign 
processes of $C_1$ to keys of $F_1$ in a  one-to-one fashion. Let $f_1$ be this map.
Then, the probability that a key
is claimed before any process under this new scheme is no lower than in a run without crashes. 
We then set $C_2$ to be the set of processes of $P_2$ that crashed before
obtaining a response for their invocation of {\tt selectWinner} in $I_2$, union
$C_1 - f_1(C_1)$. Again, after processes in $P_2$ randomly select keys in $I_2$,
assign  processes of $C_2$  to keys of $F_2$ by means of a one-to-one function $f_2$. Thus, again,
the probability that a key in $I_2$ is claimed  is higher than in a run without crashes.
And so on.
Thus, we have the following.

\begin{lemma}
The expected number of invocations of {\tt selectWinner} per process in protocol {\tt squeeze}
is $O(\log^2 n)$.
\end{lemma}

It remains to show how to implement {\tt selectWinner} in a wait-free manner
in polynomial time.
We will assume the availability of objects of type {\tt consensus(i,b)} where $1 \le i \le n$
and $b \in \{0,1\}$. Each invoking process $p$ will perform the invocation using the two
parameters $i$ and $b$; the object response will be the same to all processes 
and will be {\em ``The consensus value  for $i$ is $v$''} where $v$ is one of the bits
$b$ that were proposed. By Theorem~\ref{sym-cons-th}
the availability of consensus objects can be assumed without loss of generality. 
Assuming them will simplify the presentation.
The protocol for {\tt selectWinner}, shown in Figure~\ref{fig:select}, is as follows.
Each process $p$ generates a bit $b_1^p$ at random and invokes {\tt consensus(1,$b_1^p$)}.
Let $v_1$ be the response of the consensus object. If $b_1^p \neq v_1$ then $p$ is a {\em loser} and 
exits the protocol. Otherwise, $p$ is still in the game. Now the problem is to ascertain whether
$p$ is alone, in that case it is the {\em winner}, or if there are other processes still in the game. 
To this end, each remaining process scans the array $W[1,i]$, for $1 \le i \le n$, that is initialized
to all 0's. If $W[1,i]$ contains a 1 then $p$ declares itself a {\em loser} and exits;
otherwise it writes a 1 in its private position $W[1,p]$ and scans $W[1,-]$ again.
If $W[1,-]$ contains a single 1, namely $W[1,p]$ then $p$ declares itself the {\em winner}
and grabs the key, otherwise it continues the game that is, it generates a second bit $b_2^p$ 
at random, invokes {\tt consensus(2,$b_2^p$)}, and so on.
The following observations  and lemma establish the correctness of the protocol. 

\begin{observation}
If $p$ declares itself the winner then it is the only process to do so.
\end{observation}

\begin{observation}
There is always a process that declares itself the winner. 
\end{observation}

\begin{lemma}\label{lemma:selWinIter}
With probability
$1-o(1)$, every process $p$ generates $O(\log n)$ many random bits $b_i^p$
and the number of bit operations per process is $O(n \log n)$.
\end{lemma} 
\begin{proof}
We refer to an iteration of a repeat loop of protocol {\tt selectWinner} as a {\em round},
i.e. the round $i$ refers to the
set of iterations of the repeat loop in which the participating processes 
toss their private coin for the $i$th time. 
We assume pessimistically that the consensus object of protocol {\tt selectWinner} is under the
control of the strong adversary, subject to the following rules. Denoting with 
$1, \ldots, k$ the processes that perform the $i$th coin toss, 
If $<b_i^1 = b_i^2 = \ldots = b_i^k = 0$ or 
$<b_i^1 = b_i^2 = \ldots = b_i^k = 1$ then the adversary can respond with consensus value
0 or 1, respectively. Otherwise the adversary can respond with any value.

The goal of the adversary is to maximize the number of rounds. Therefore
its best policy is to return the consensus value that eliminate the smallest number
of processes, i.e. the best strategy is to return the majority bit. 
The probability that the $i$th outcome of process $p$ is the majority value
depends on the number of processes, but it is easily seen to be maximized when there
are two processes. 
The probability that the minority value is the outcome of at least $1/4$
of the processes depends on the number of processes, but it is monotonically decreasing.
Therefore the smallest value is $1/4$, when just 2 processes are involved. 

We call a run {\em successful} if the minority value is the outcome of at least $1/4$
of the processes. Then, $\log_{4/3} n$ many successful rounds suffice to select a winner.
A straightforward application of the Chernoff-Hoeffding bounds
show that with probability $1-o(1)$ at least $\log_{4/3} n$ rounds out of $8 \log_{4/3} n$
many will be successful. 

Since every iteration of {\tt selectWinner} costs $O(n)$ steps, the claim follows.
\end{proof}

\begin{theorem}
Protocol {\tt squeeze} is a Las Vegas, wait-free naiming protocol for asymmetric memory 
whose running time is $O(n^2 \log n)$ with probability $1 - o(1)$.
\end{theorem}

\noindent
{\bf Remark 1:} Protocol {\tt squeeze} is also a good renaming protocol.
Instead of the random bits, each process can use the bits of its own IDs starting, say, from the
left hand side.
Since the ID's are all different the above scheme will always select a unique winner
within $O(|ID|)$ invocation of consensus.
\bigskip

\noindent
{\bf Remark 2:} The only part of protocol {\tt squeeze} that actually uses the
memory is protocol {\tt selectWinner}. In view of Proposition~\ref{fact:namImp} this task
must be impossible with symmetric memory, even if randomness and consensus are available.
Thus, this is another task for which, strictly speaking, Herlihy's result does
not hold and it is another 
example of something that cannot be accomplished by the power of randomization alone.


\begin{figure}
\begin{center}
\small
\hrule
\tt
\begin{tabular}{l}
\\
protocol squeeze(): key; \\[.2cm]
begin \\
\ \ for i := 1 to $\ell$ do begin \\
\ \ \ \ k := random key in interval $I_i$; \\
\ \ \ \ if selectWinner(k) = {\em ``You own key $k$!''} then return(k); \\
\ \ \ \ end; \\
\ \ for k := n - $s_{\ell}$ to n do \hfill \{try key in $I_{\ell}$ one by one\} \\
\ \ \ \ if selectWinner(k) = {\em ``You own key $k$!''} then return(k); \\
\ \ return(simpleButExpensive()) \hfill \{back up procedure\} \\
end \\
\\
\end{tabular}
\hrule
\end{center}
\caption{Protocol squeeze}\label{fig:squeeze}
\end{figure}

\begin{figure}
\begin{center}
\small
\tt
\hrule
\begin{tabular}{l}
\\
protocol selectWinner(i: key): outcome; \\
\\
myReg := ``private register of executing process''; \\
attempt := 1; \\
repeat \\
\ \ b := random bit; \\
\ \ if (b = consensus(i, b)) then begin \\
\ \ \ \ scan W[attempt,j] for $1 \le j \le n$; \\
\ \ \ \ if  (W[attempt,j] = 0, for all j) then begin \\
\ \ \ \ \ \ W[attempt,myReg] := 1; \\
\ \ \ \ \ \ scan W[attempt,j] for $1 \le j \le n$; \\
\ \ \ \ \ \ if  (W[attempt,j] = 0, for all j <> myReg) then return(i); \hfill \{key
 is grabbed!\} \\
 \ \ \ \ \ \ else attempt := attempt + 1;  \hfill \{keep trying\} \\
 \ \ \ \ else return({\em ``Sorry, look for another key.''}); \\
 end repeat \\
 \\
 \end{tabular}
 \hrule
 \end{center}
 \caption{Protocol selectWinner}\label{fig:select}
 \end{figure}

\end{document}